\documentclass[aps,pra,twocolumn,superscriptaddress]{revtex4-1}
\usepackage[utf8]{inputenc} 
\usepackage[T1]{fontenc} 
\usepackage[english]{babel}
\usepackage[autostyle=true]{csquotes} 
\usepackage[dvipsnames]{xcolor}
\usepackage{graphicx}
\usepackage{subcaption}
\usepackage{array}
\usepackage{amsmath} 
\usepackage{amssymb} 
\usepackage{amsfonts} 
\usepackage{amsbsy}
\usepackage{mathrsfs} 
\usepackage{dsfont} 
\usepackage{dcolumn} 
\usepackage{upgreek} 
\usepackage{bm} 
\usepackage{bbm}
\usepackage{placeins} 
\usepackage{physics} 
\usepackage[compat=1.1.0]{tikz-feynman}
\usepackage{booktabs} 
\usepackage[colorlinks,linkcolor=blue,citecolor=blue,urlcolor=blue]{hyperref}
\usepackage[most]{tcolorbox}
\usepackage{float}
\usepackage[sort&compress]{natbib}
\usepackage[dvips]{epsfig}
\usepackage{hhline}
\usepackage{multirow}
\usepackage[normalem]{ulem}
\usepackage{xspace}
\newcommand{\remove}[1]{}

\def\ppe        {P\&p\ }

\def\trarpes{{\it tr}ARPES\xspace}
\renewcommand{\[}{\left[}
\renewcommand{\]}{\right]}
\renewcommand{\(}{\left(}
\renewcommand{\)}{\right)}

\def\rar        {\rightarrow}

\newcommand{\eq}[1]{\begin{align}#1\end{align}}

\newcommand{\eqg}[1]{\begin{gather}#1\end{gather}}

\newcommand{\seq}[1]{\begin{subequations}#1\end{subequations}}

\newcommand{\seql}[2]{\begin{subequations}\label{#1}#2\end{subequations}}
\newcommand{\mll}[2]{\begin{multline}\label{#1}#2\end{multline}}
\newcommand{\eql}[2]{\begin{align}\label{#1}#2\end{align}}
\newcommand{\eqgl}[2]{\seq{\label{#1}\begin{gather}#2\end{gather}}}

\newcommand{\stkout}[1]{\ifmmode\text{\sout{\ensuremath{#1}}}\else\sout{#1}\fi}

\newcommand{\lab}[1]{\label{#1}}

\newcommand{\ul}[1]{\underline{#1}}

\newcommand{\e}[1]{Eq.~\eqref{#1}}

\newcommand{\elab}[2]{Eq.(\ref{#1}#2)}

\newcommand{\fig}[1]{Fig.\ref{#1}}
\newcommand{\figlab}[2]{Fig.\ref{#1}#2}

\newcommand{\h}[1]{\hat{#1}}

\newcommand{\ocite}[1]{Ref.\cite{#1}}
\newcommand{\p}{\prime}           
\newcommand\vk{{\mathbf{k}}}
\newcommand\vq{{\mathbf{q}}}

\newcommand\hc{{\hat{c}}}
\newcommand\hcd{{\hat{c}^\dagger}}

\newcommand\hx{{\hat{x}}}
\newcommand\hp{{\hat{p}}}
\newcommand\hH{{\hat{H}}}

\newcommand\hrho{{\hat{\rho}}}

\newcommand\upi{\mathord{\mathrm{i}}}

\newcommand\ik{{i\vk}}

\newcommand\HF{\text{HF}}
\newcommand\Eh{\text{Eh}}

\newcommand\urho{{\underline{\rho}}}

\newcommand\uDrho{{\Delta\underline{\rho}}}

\newcommand\ug{\underline{g}}

\def\gC         {\Gamma}

\def\go         {\omega}

\def\gS         {\Sigma}

\def\zero	{{\mathbf 0}}

\def\QQ		{{\mathbf Q}}

\def\kk		{{\mathbf k}}
\def\qq		{{\mathbf q}}

\newcommand{\cnrism} {Istituto di Struttura della Materia and Division of Ultrafast Processes in Materials (FLASHit) of the National Research Council, via Salaria Km 29.3, I-00016 Monterotondo Stazione, Italy}

\newcommand{\mysec}[1]{{\em #1}}

\begin{document}

\title{
Excitons in WSe$_2$ time--resolved ARPES: particle or oscillation?
}
\author{Kai Wu}
\affiliation{\cnrism}

\author{Michele Puppin}
\affiliation{Laboratory of Ultrafast Spectroscopy (LSU) and Lausanne Centre for Ultrafast Science (LACUS), École Polytechnique Fédérale de Lausanne (EPFL), CH-1015 Lausanne, Switzerland}

\author{Andrea Marini}
\affiliation{\cnrism}
\begin{abstract}
The time--resolved angle--resolved photoemission spectra of WSe$_2$, a paradigmatic transition metal dichalcogenide, are dominated by a transient signal that,
after being initially observed in the gap at the $K$ valley, scatters, on an ultra--fast time scale of $\sim\,30$\,fs, to the $\gS$ valley.
In this work we question the common interpretation of the experimental dynamics in terms of a massive bound electron--hole exciton
that scatters with phonons and behaves as a quasi--particle.
By using a combined theoretical and experimental investigation, we demonstrate that the observed dynamics
can be interpreted as the photo--induced transition from direct to indirect excitonic--insulating order.
The features that appear in the experimental spectrum correspond to
single--particle levels renormalized by the excitonic spontaneous polarization.

\end{abstract}
\date{\today}
\maketitle


\mysec{Introduction.}
A direct excitonic insulator\,(DEI) is a phase in which Coulomb attraction drives a spontaneous interband coherence in small direct--gap semiconductors or
weakly overlapping semimetals\cite{JeromeRiceKohn1967}. The DEI phase is characterized by an excitonic order parameter $\Delta_{\mathbf{k}}$, which measures the
level of hybridization between valence and conduction bands, responsible for the induced electronic polarization. 

At equilibrium the DEI phase is characterized by: gap opening, dissipationless Coulomb drag, interlayer coherence, spectral-weight transfer and characteristic band
hybridization, which can be accessed by photoemission and optical probes~\cite{Butov2002,Wang2019}.  In particular, angle--resolved photoemission
spectra\,(ARPES) measurements on Ta$_2$NiSe$_5$
reported a pronounced flattening and hybridization near the valence-band edge across the transition~\cite{Wakisaka2009}.  

In their original paper~\cite{JeromeRiceKohn1967}, J\'erome, Rice and Kohn predicted that, in the presence of an indirect gap, the system can undergo a
transition to an indirect excitonic insulating\,(IEI) phase where, in addition to the spontaneous appearance of a macroscopic polarization, also the
symmetry of the ground state is broken by the momentum, $\QQ$, corresponding to the difference between
the positions of the conduction-band minima\,(CBM) and valence-band maxima\,(VBM). The IEI is characterized by a finite momentum order parameter,
$\Delta_{\mathbf{k}}\(\QQ\)$.

The IEI physics is particularly rich in layered materials, like  transition metal dichalcogenides\,(TMDs), where the van der Waals interaction binds the layers
allowing fine tuning of the underlying electronic structure by manipulating the layers geometry.  The transition between DEI and IEI has been
proposed~\cite{Merkl2019}, for example, to interpret near--infrared pump and mid--infrared probe spectroscopy of van der Waals heterostructures. The IEI, like
the DEI, is characterized by a Bose condensation regime\cite{Wu2015}.

Out--of--equilibrium physics, and in particular time--resolved ARPES\,(\trarpes) experiments, may provide a direct way to create and observe a transient DEI
phase~\cite{Perfetto2019,Murakami2017,Golez2020}. The possibility of using a laser field to drive the formation of an excitonic order parameter was already
discussed in 1988 by Schmitt--Rink et al.~\cite{Schmitt-Rink1988} as a manifestation of a time--dependent Stark effect. Schmitt--Rink showed that the 
macroscopic polarization appearing in the photo--induced DEI phase 
renormalizes the single--particle levels.
The authors also observed a close analogy between the equation of motion for the DEI order parameter and the physics of excitonic insulators, superconductors and Bose
condensed systems.  The same analogies were confirmed in \ocite{Perfetto2019}, where the observed photo--induced band modifications were shown to induce
an excitonic level in the direct gap, in agreement with \trarpes experiments. 

In the case of TMDs, \trarpes\, experiments have made it possible to observe,
with unprecedented precision, the valley dynamics of the excitonic features~\cite{dong2021,Madeo2020,Trovatello2020}.
In particular, in \ocite{dong2021}, an initial
signal inside the direct gap at $K$ that,   on an ultra--fast time scale of $\sim\,30$\,fs, scatters to the $\gS$ valley. This signal has been mostly 
interpreted~\cite{Rustagi2018,Steinhoff2017,Christiansen2019,Katzer2023} as a real, bound electron--hole pair (i.e. an exciton) that scatters with phonons as a real
quasiparticle. 

There are, however, some aspects that question such a quasiparticle representation.  
A crucial property of the photo--induced DEI is that the order parameter is non--linear in the external perturbation. This has been demonstrated by
\ocite{Schmitt-Rink1988,Perfetto2019}. This rules out any description in terms of linear optical\footnote{For optical excitons we refer to the poles of the
macroscopic dielectric function~\cite{Onida2001}, commonly observed in absorption experiments} excitons and requires a non perturbative treatment where the
interpretation of the photo--excited dynamics in terms of electron--hole pairs is questionable.
Moreover, the experimental evaluation of the intrinsic linewidth of the bound optical WSe$_2$
exciton~\cite{Moody2015} gives, for the total width, $\gC^{exc}_K\lesssim\,5$meV, which corresponds to a $\tau^{exc}_K\gtrsim\,130$\,fs. This lifetime, however, corresponds to the
excitonic scattering to all possible final states, including the $\gS$ valley. The final $K\rar\gS$ estimate of the excitonic scattering time is, therefore,
incompatible with the experimental results\cite{dong2021}.



\mysec{In this work}
We propose that a {\em non--linear} direct-to-indirect excitonic insulating phase transition explains the fast scattering observed experimentally in WSe$_2$.
We demonstrate that the observed \trarpes spectrum is an experimental realization of this transition.
The proposed interpretation is valid beyond the linear regime and, thus, does not require the introduction of a quasiparticle excitonic picture.
Physically, we interpret the inter--valley scattering in terms of a spontaneous polarization whose order parameter adiabatically follows the
slower carriers that scatter from $K$ to $\gS$.
The \trarpes signal then reflects the dynamical Stark correction of the single-particle levels induced by
the direct and indirect excitonic polarizations.
In agreement with the prediction of J\'erome, Rice and Kohn \cite{JeromeRiceKohn1967}, we reveal 
that the indirect spontaneous polarization causes the breakdown of lattice periodicity,
giving rise to a lattice order parameter that oscillates together with the electronic one.
\begin{widetext}
Our theory leads to excellent agreement with the experimental results as shown in \fig{fig:1}, providing an intuitive explanation of the basic
mechanism that drives the excitonic features in the \trarpes of WSe$_2$.
\begin{figure}[h]
	\centering
	\begin{tikzpicture}
		\node[anchor=center] (figone) at (0,0) {\includegraphics[width=\columnwidth]{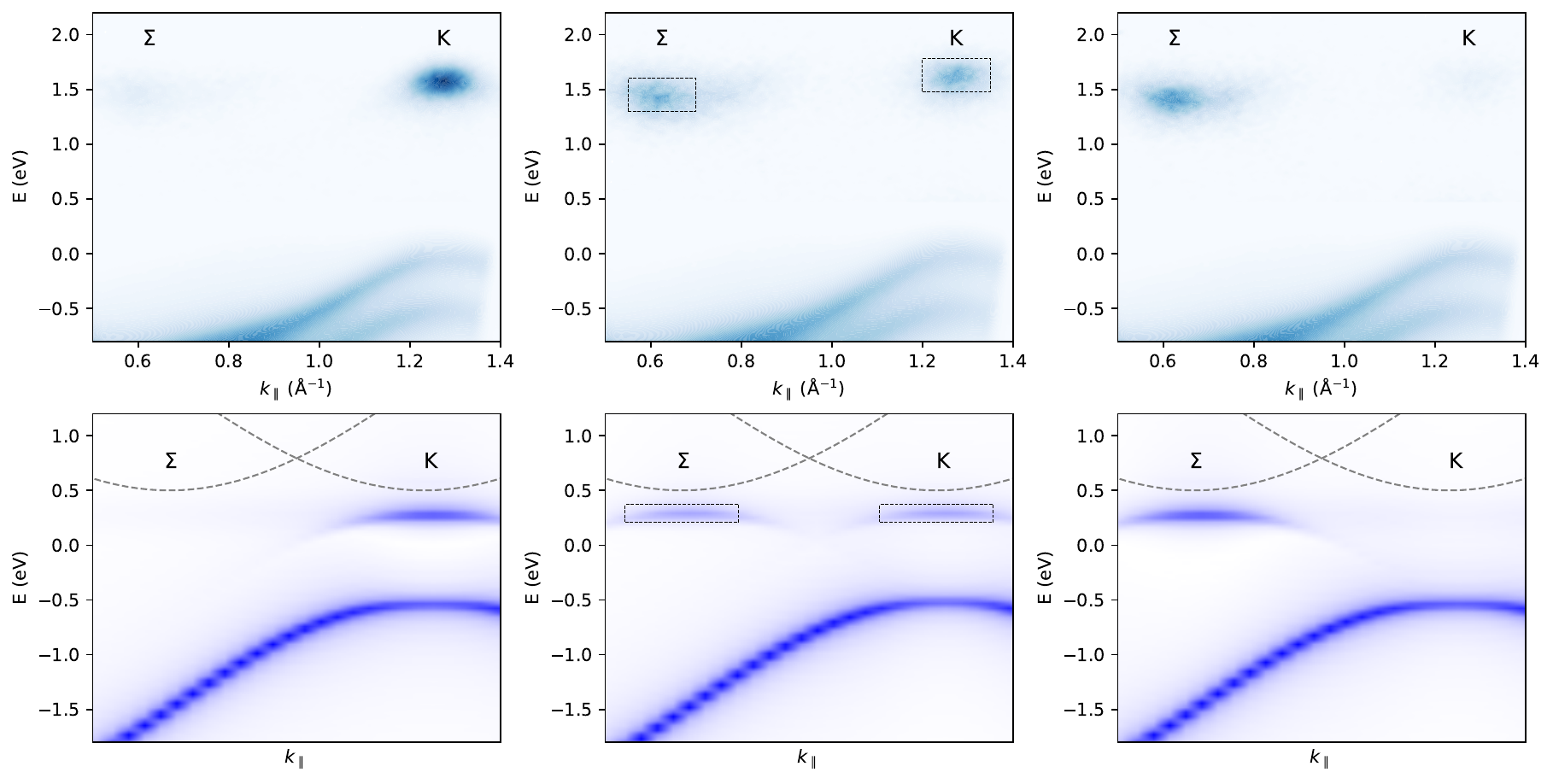}};
		\node[anchor=center] at (-8.7,4.3) {(a)};
		\node[anchor=center] at (-2.7,4.3) {(b)};
		\node[anchor=center] at (3.2,4.3) {(c)};
		\node[anchor=center] at (-8.7,-.2) {(d)};
		\node[anchor=center] at (-2.7,-.2) {(e)};
		\node[anchor=center] at (3.2,-.2) {(f)};
	\end{tikzpicture}
	\caption{\footnotesize{Time-resolved ARPES intensity in WSe$_2$: experiment versus theory.
			Panels (a)--(c): experimental photoemission maps $I(E,k_{\parallel})$ along the $\Sigma$ to $K$
			direction at pump--probe delays of $-40$, $-10$, and $200$ fs, respectively.
			Panels (d)--(f): simulated spectra at the same delays.
			Dashed curves indicate the underlying valley dispersions.
			Dashed boxes in panels (b) and (e) indicate the momentum--energy windows used to evaluate the
			integrated signal, compared with the experiment in \fig{fig:exc_curve}}}
\lab{fig:1}
\end{figure}
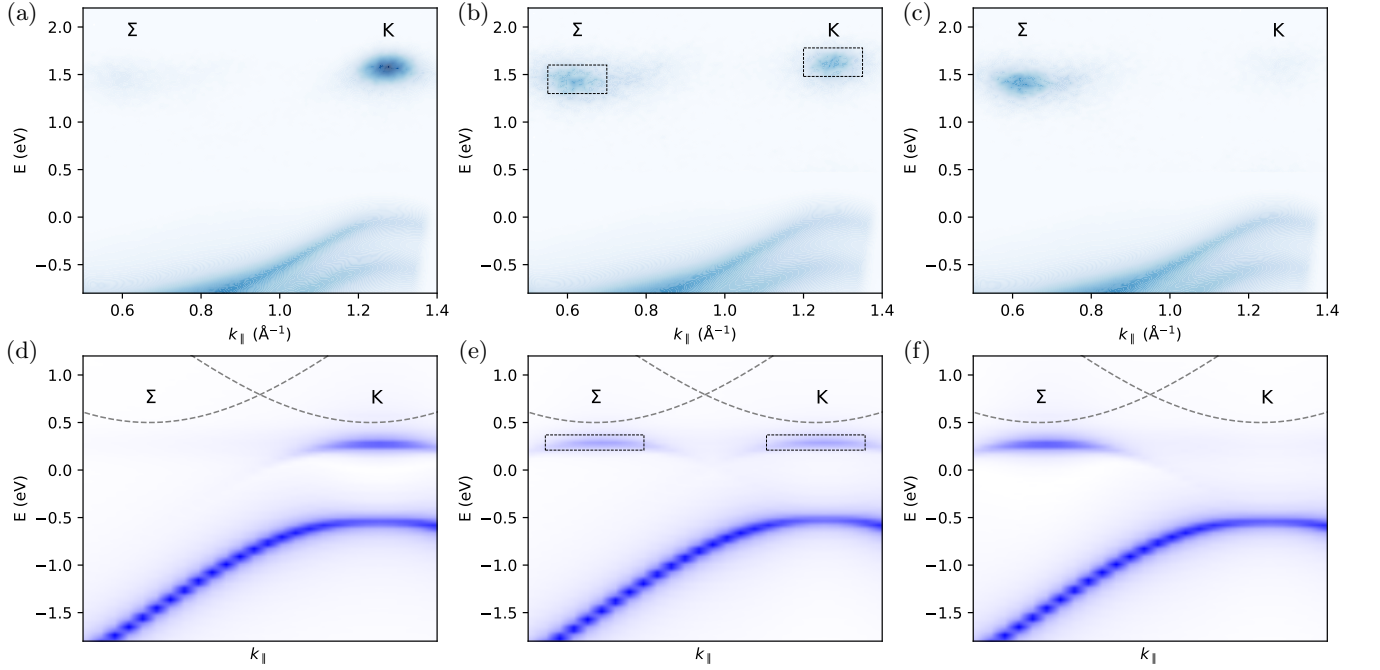
\end{widetext}

\mysec{Active regions and Hamiltonian.}
To capture the mechanism underlying the \trarpes response of WSe$_2$, we introduce a two--band time--dependent Hamiltonian $\hH\(T\)$
following the strategy used previously in~\ocite{Murakami2017,Golez2020}.
The Hamiltonian describes a spinless electronic system coupled to an optical phonon mode and excited by an
external pump laser field $P\(T\)$:
\eql{eq:H_total_c}{
 \hH\(T\)=\hH_0+\hH_{pump}\(T\)+\hH_{e-p}+\hH_{e-e},
}
with
\seql{eq:H_comp}{
\eqg{
 \hH_0=\sum_{i\vk}\epsilon_{i\vk}\hcd_\ik\hc_\ik+\frac{1}{2}\sum_{\vq}\omega_{\vq}\Big(\hx_{\vq}\hx_{-\vq}+\hp_{\vq}\hp_{-\vq}\Big),\\
 \hH_{pump}\(T\)=P\(T\)\sum_\vk\(\hcd_{2\vk}\hc_{1\vk}+h.c.\),\\
 \hH_{e-e}=\frac{1}{N}\sum_{\vk \vk^\p\vq}U\hcd_{1\vk+\vq}\hcd_{2\vk^\p-\vq}\hc_{2\vk^\p}\hc_{1\vk}-U\sum_\vk\hcd_{2\vk}\hc_{2\vk},\\
 \hH_{e-p}=\frac{1}{\sqrt{N}}\sum_{\vk\vq ij}\tr\!\left[g_{ij\vk}(\vq)\,\Delta \hrho_{ji\vk+\vq}(-\vq)\right]\hx_{\vq}.
}
}
$\hc_{i\vk}$ annihilates an electron with crystal momentum $\vk$ in band $i$ ($i=1$ valence, $i=2$ conduction), $N$ is the total number of
$\vk$-points.  The one--dimensional band dispersions are chosen as $\epsilon_{1\kk}=\cos(k)-1-E_g/2$, $\epsilon_{2\kk}=1-\cos(k-\QQ)+E_g/2-U$.
$\hH$ depends on $\QQ$ via $\epsilon_{n\kk}$, where $\QQ$ is the momentum separation between the conduction-band minimum\,(CBM) at $\gS$ and the valence-band
maximum\,(VBM) at $K$. This allows us to describe, at the same time, the two valleys.  We model the electron–hole attraction by a $\vq$-independent constant $U$,
which can be viewed as an average of the statically screened interaction.

The pump laser field is taken to be 
$P\(T\) = A \exp\!\left[-\frac{(T-T_0)^2}{2\sigma^2}\right]$.
The operators $\hx_{\vq}$ and $\hp_{\vq}$ denote the phonon displacement and momentum, respectively.  We define $\hrho_{ij\kk}(\qq)=
\hcd_{j\kk}\hc_{i\kk+\qq}$, and $\Delta \hrho_{ij\kk}(\qq)=\hrho_{ij\kk}(\qq)-\rho_{ij\kk}(\qq)^{\text{ref}}$ with respect 
to a reference density matrix $\rho^{\text{ref}}$, which serves as the
expansion point for the harmonic approximation used to introduce phonons~\cite{Marini2015,Stefanucci2023}.  For simplicity, we consider a single optical phonon
branch and retain only interband electron--phonon coupling, so that $g_{ij\vk}(\vq)$ is nonzero only for $i\neq j$.

\mysec{Time--scale separation.}
\trarpes is resolved as a function of the pump--probe delay $T$ (measured with respect to $T_0$) and the photo--emitted energy $\go$. 
$\go$ is obtained from the Fourier transform with respect to the time $\tau$ that describes the system dynamics after photo--excitation. 

Both the IEI and DEI are variational states and, thus, correspond to an instantaneous solution of the self--consistent problem.
This suggests separating the dynamics into two phases: in the first phase\,(IEI--DEI phase), after photo--excitation, the carriers evolve on
the macroscopic time $T$. As suggested by \ocite{Perfetto2019}, at each time $T$ the system is driven into a coherent excitonic insulating phase. In the
second phase (probe phase), at each observed time the system evolves without the conduction electrons, removed by the probe pulse. This
second-phase dynamics is described by the time $\tau$.

We refer to the time separation in the IEI--DEI phase as non--equilibrium Williams--Lax approach\,(NE--WLA).
The WLA~\cite{williams1951theoretical,lax1952franck} is an approach to study the combined electron--nuclei
dynamics.  
In the NE--WLA the carriers evolve slowly, akin to the lattice in the standard WLA approach, whereas the
excitonic phase instantaneously adapts to the carriers.
In practice, we first determine the $T$ valley populations from a
Pauli master equation, then construct the corresponding excitonic phase from the variational solution of a self--consistent Hamiltonian that represents
a picture of the excited system at time $T$.  
Finally, the evolution on the probe time axis
$\tau$\,(conjugate to the photo--emitted energy $\go$) is performed to obtain the spectral function to be compared with experiment.

\mysec{Macroscopic time--scale: the incoherent carriers dynamics.}
The incoherent carrier dynamics induced by \e{eq:H_total_c} is described at the level of a Markovian
two--state Pauli master equation, motivated by the Generalized Baym--Kadanoff ansatz~\cite{Leeuwen2013}
and by the Markov approximation~\cite{Marini2013}.
The pump injects carriers into the $K$ valley through a source term $P\(T\)$. At this point the dynamics is described as scattering
from $K$ to $\gS$ with an experimental lifetime $\tau_{K\rar\gS}=19$~fs taken from \ocite{Selig2016}.
In addition the $\gS$ electrons can scatter back to $K$ or to other states. This scattering has lifetime $\tau_{\gS}$.
Note that $\tau_{K\rar\gS}$ is much shorter than $\tau^{exc}_{K\rar\gS}$ as it involves free carriers, not bound
in an electron--hole pair.

The equations of motion read
\eqgl{eq:carr}{
 \frac{d}{dT} n_K\(T\)=P\(T\)-\frac{1}{\tau_{K\rar\gS}}\,n_K\(T\)+\frac{1}{\tau_{\gS \rar K}}\,n_{\gS}\(T\),\\
 \frac{d}{dT} n_{\gS}\(T\)=\frac{1}{\tau_{K\rar\gS}}\,n_K\(T\)-\frac{1}{\tau_{\gS}}\,n_{\gS}\(T\).
}
In \e{eq:carr}, $\tau_{\gS\rar K}$ denotes the $\gS\rar K$ contribution to the total lifetime $\tau_\gS$ \footnote{$\tau_{\gS\rar K}>\tau_{\gS}$ is used
as a parameter to tune the correct balance between the $K$ and $\gS$ carrier populations}.


\mysec{Self--consistent Hamiltonian and coherent EI phase.}
At each macroscopic time $T$, following \ocite{Murakami2017,Golez2020,Perfetto2019}, we use the $n_K\(T\)$ and $n_\gS\(T\)$ extracted from \e{eq:carr}
to construct an instantaneous Hamiltonian at the mean--field level:
\mll{eq:scf_Hm}{
 \ul{H}[\delta\mu(T),\rho(T),x(T)]=\\
 \ul{h}[\delta\mu(T)]+\ul{\Sigma}^\HF[\rho\(T\)]+\ul{\Sigma}^\Eh[x\(T\)].
}
Here $\rho\(T\)$ and $x\(T\)$ are the electronic density matrix and lattice displacement, and $\Sigma^\HF$ and $\Sigma^\Eh$ denote the Hartree--Fock and
Ehrenfest self--energies, respectively.  
In \e{eq:scf_Hm} all terms (denoted as $\ul{O}$) are matrices in the single--particle basis and can be calculated at zero and finite momentum.
Following \ocite{Perfetto2019}, the action of the pump field is incorporated through an equivalent time--dependent
shift of the single--particle levels
\mll{eq:mu}{
 h_{ij\vk}(\vq)[\delta\mu(T)]=\bra{i \kk}\h{h}\[\delta\mu\(T\)\]\ket{j\kk-\qq}=\\
\delta_{\vq0}\delta_{ij}\big(\epsilon_{i\vk}-\delta_{i2}U_0+(\delta_{i1}-\delta_{i2})\delta\mu(T)/2\big). 
}
The values of $\delta\mu\(T\)$, $\rho\(T\)$, and $x\(T\)$ are determined self--consistently by solving the eigenvalue problem
\mll{eq:scf_equation}{
	\sum_{m\vq} H_{im\,\vk-\vq}(\vq)[\delta\mu(T),\rho(T),x(T)]\,\psi^\lambda_{m\,\vk-\vq}\(T\)\\
	=e^\lambda\(T\)\,\psi^\lambda_{i\vk}\(T\),
}
which yields the quasiparticle energies $e^\lambda\(T\)$ and eigenvectors $\psi^\lambda_\ik\(T\)$.
The initial value of $\delta\mu\(T\)$ in the self--consistent cycle is chosen to reproduce the carrier populations $n_K\(T\)$ and $n_\gS\(T\)$ evaluated from
\e{eq:carr}. 

During the iteration cycle, the density matrix and lattice displacement are updated according to
\eqgl{eq:scf_update}{
	\rho_{ij\vk}(\qq,T)=\sum_\lambda f(e^\lambda)\psi^\lambda_{i\vk+\vq}(T)\psi^{\lambda*}_{j\vk}(T),\label{eq:scf_update_a}\\
	x_{\vq}\(T\)=-\frac{1}{\sqrt{N}\,\omega_{\vq}}\(\sum_{\vk}\tr\!\[\ug_{\vk}(-\qq)\uDrho_{\vk-\vq}\(\qq,T\)\]\),\label{eq:scf_update_b}
}
where $f\(e^\lambda\)$ is the Fermi--Dirac distribution.
\e{eq:scf_update_b} is equivalent to enforcing the stationarity condition ${\partial E\(T\)}/{\partial
x_{\vq}\(T\)}=0$ which corresponds to zero forces acting on the lattice. Physically this corresponds to an adiabatic lattice dynamics that instantaneously adapts
the atoms to the slowly varying electronic dynamics.


\mysec{Excitonic phase order parameters.}
\e{eq:scf_update} defines two order parameters
\eqgl{eq:OP}{
\Delta_{\vq}\(T\)=\sqrt{\frac{1}{N}\sum_\vk\left|\rho_{21\vk}(\qq,T)\right|^2},\\
X_\vq\(T\)=|x_{\vq}\(T\)|.
}
$\Delta_{\vq}\(T\)$ is the time--dependent electronic order parameter that extends the one defined in \ocite{Perfetto2019,Murakami2017,Golez2020}
to time and momentum. $X_\vq\(T\)$ is an additional lattice order
parameter connected to the lattice symmetry periodicity breakdown. The $\qq$ and $T$ dependences imply that 
the two order parameters can move in time between the $K$ and $\gS$ valleys.
Here we use a valley shorthand for the transferred momentum:
the direct component is denoted by
$\Delta_K(T)\equiv\Delta_{\vq=\zero}(T)$, and $X_K(T)\equiv X_{\vq=\zero}(T)$,
while the indirect electronic and lattice components are denoted by
$\Delta_\Sigma(T)\equiv\Delta_{\vq=\QQ_\gS-\QQ_K}(T)$
and
$X_{-\Sigma}(T)\equiv X_{\vq=\QQ_K-\QQ_\gS}(T)$.

\begin{figure}[t]
	\centering
	\begin{tikzpicture}
		\node[anchor=center] at (0,0) {\includegraphics[width=\columnwidth]{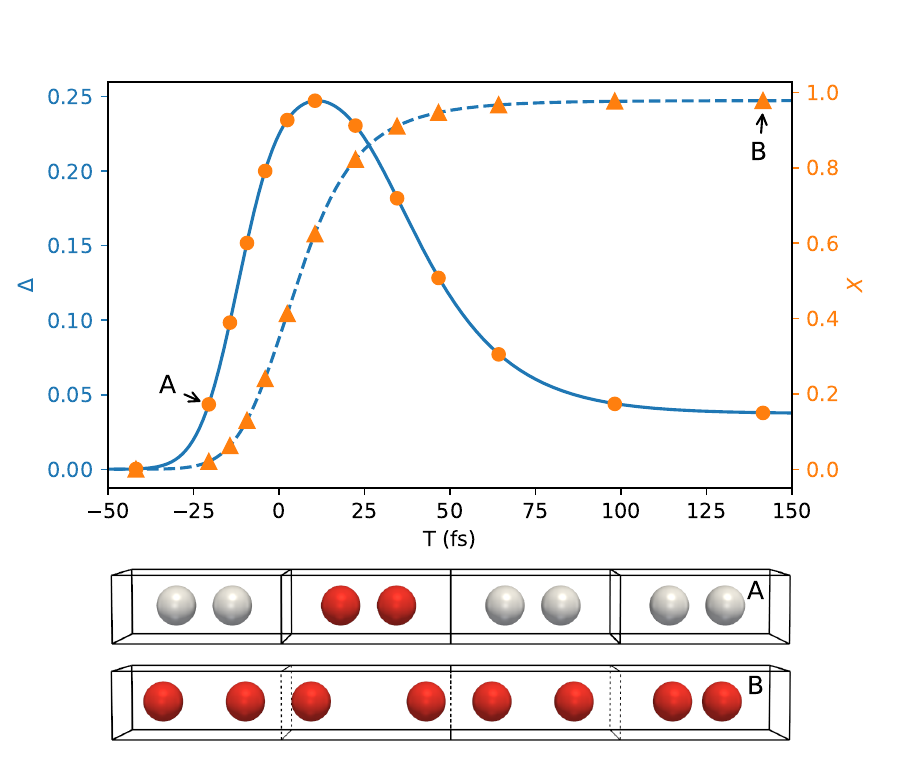}};
	\end{tikzpicture}
	\caption{
		\footnotesize{
			Macroscopic-time evolution of the electronic and lattice order parameters.
			The blue solid line denotes $\Delta_K$, the blue dashed line denotes $\Delta_\Sigma$,
			the orange circles denote $X_K$, and the orange triangles denote $X_{-\Sigma}$.
			The lower panels show schematic lattice configurations at the representative times A and B,
			before and after symmetry breaking, respectively.
	}}
	\label{fig:order_parameter_T}
\end{figure}

Indeed the simulations reveal that at small $T$, $\Delta_{\vq=\zero}\(T\)$ dominates. In this regime, see \fig{fig:1},
the \trarpes is dominated by a clear signal in the $K$ valley. The corresponding lattice order parameter is associated with a
symmetric lattice dynamics. When $T$ grows and carriers move to the $\gS$ valley, the electronic and lattice finite-momentum
order parameters increase, pointing to a coherence transfer between the two valleys.

This is represented in \fig{fig:order_parameter_T}. We consider two times, $T_{A}$ and $T_B$. At $T=T_A$
the lattice symmetry is not broken, while at $T=T_B$ the atomic displacement is not periodic and 
the corresponding lattice order parameter $X_{-\gS}\(T_B\)\neq 0$. It is remarkable to observe that
at $T=T_B$ the electronic and lattice order parameters are populated at opposite momenta, reflecting overall momentum conservation (the system is isolated
and the external laser pump does not transfer momentum to the system).

\mysec{Photoemission and time--resolved spectral function.}
\begin{figure}[t]
	\centering
	\begin{tikzpicture}
		\node[anchor=center] at (0,0) {\includegraphics[width=.9\columnwidth]{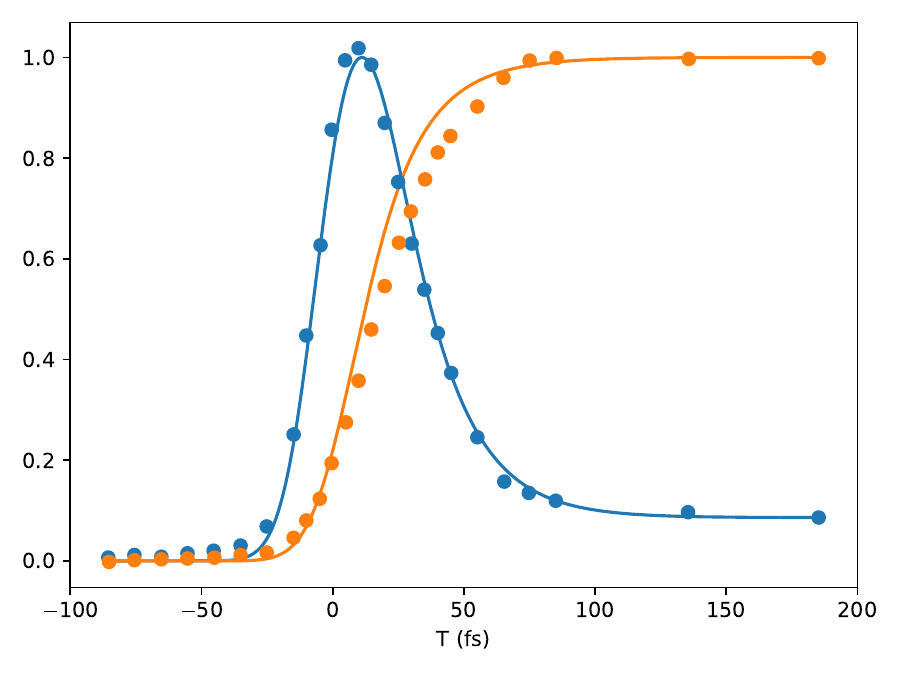}};
	\end{tikzpicture}
\caption{\footnotesize{Simulated time--delay dependence of the excitonic signals at the $K$ (blue line) and $\gS$ (orange line) valleys. Dots are the experimental data.
	}}
	\label{fig:exc_curve}
\end{figure}
The macroscopic time $T$ is measured with respect to the pump center and, thus, corresponds to the
\ppe delay. Here it corresponds to the time at which the probe removes the conduction electron.
In order to calculate the energy-dependent \trarpes current, we follow \ocite{Perfetto2019} and introduce
the photo--emission time $\tau$ (conjugate to the measured energy $\go$).
We start
from the prepared initial state $\psi^\lambda_\ik\(T,\tau=0\)=\psi^\lambda_\ik\(T\)$ and $x_\vq\(T\)$.
Within the sudden approximation, photoemission is modeled as an effective electron--removal process
acting on the prepared state.
Accordingly, we set $\delta\mu(T)=0$ (corresponding to a sudden removal of the conduction electrons) and
propagate the microscopic dynamics
\begin{widetext}
\seql{eq:real_time}{
\eq{
\upi\frac{d}{d\tau}\psi^\lambda_\ik(T,\tau)= \sum_{m\vq}H_{im\,\vk-\vq}(\vq)[0,\urho_\vk(\vq,T,\tau),x_\vq(T,\tau)]\psi^\lambda_{m\,\vk-\vq}(T,\tau),
}
and
\eq{
\frac{d^2}{d\tau^2}x_{\vq}(T,\tau)=-\omega_{\vq}^2 x_{\vq}(T,\tau)-\frac{\omega_{\vq}}{\sqrt{N}}\sum_\vk\tr(\ug_\vk(-\qq)\uDrho_{\vk-\vq}(\qq,T,\tau)),\\
}}
\end{widetext}
where $\urho_\vk(\vq,T,\tau)$ is constructed similarly to \e{eq:scf_update_a}.
The $\tau$- and $T$-dependent eigenfunctions $\psi^\lambda_{m\,\vk-\vq}(T,\tau)$ can be used to reconstruct the order-parameter oscillations following
photo--excitation through \e{eq:scf_update}.  

The spectral function can be reconstructed from
\mll{eq:A_wigner}{
	A^<_\vk(T,\omega)=\int_0^\infty d\tau\; e^{i(\omega+i\eta)\tau}
	\sum_{i\lambda} f(e^\lambda)\\
	\times\psi^\lambda_{i\vk}(T,\tau_0+\frac{\tau}{2})\psi^{\lambda*}_{i\vk}(T,\tau_0-\frac{\tau}{2}),
}
where $\tau_0$ is chosen such that $\tau_0-\tau/2\ge 0$ over the integration range, and $\eta>0$ is a small
broadening parameter that ensures convergence of the Fourier transform.
Representative spectra at different delays are compared with the \trarpes measurements in \fig{fig:1}.

We then calculate the conduction--band excitonic intensity $I\(T\)$ at the $K$ and $\gS$ valleys by integrating the spectral weight over the
momentum--energy windows enclosing the excitonic feature, indicated by the dashed boxes in \fig{fig:1}.
The resulting time-dependent spectral intensity is shown in \figlab{fig:exc_curve} and compared with the experimental data.
We see that the agreement with the experimental intensity of the \trarpes signal is excellent, showing the robustness and accuracy of the proposed interpretation.

\mysec{Coherent phonon oscillations.}
The \trarpes provides access to the electronic degrees of freedom and, as explained above, the observed spectrum is the Fourier transform of
the $\tau$-dependent spectral function. By comparing \e{eq:A_wigner} and \elab{eq:scf_update}{1} it follows that the spectral features
are due to the electronic order parameter oscillation in $\tau$.

\begin{figure}[t]
	\centering
	\begin{tikzpicture}
		\node[anchor=center] at (0,0) {\includegraphics[width=0.9\columnwidth]{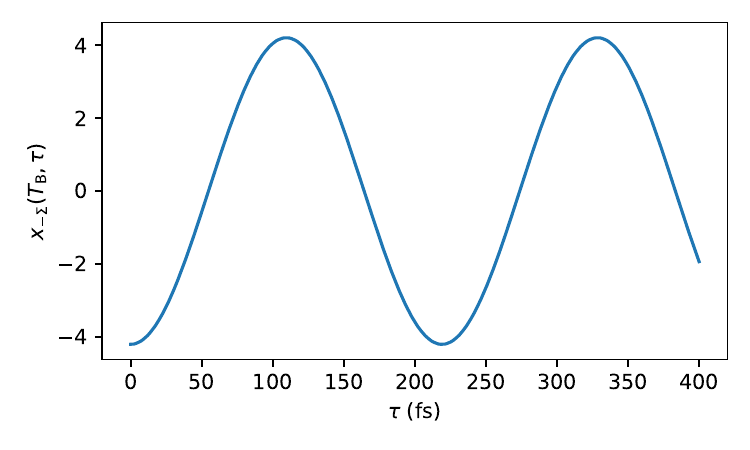}};
	\end{tikzpicture}
	\caption{\footnotesize{
	  Time evolution of the phonon displacement $x_{-\Sigma}(T_\mathrm{B},\tau)$ after probe photo--excitation, where $\tau$ is measured from $T_\mathrm{B}$.
	}}
	\label{fig:phonon_osc}
\end{figure}

Clearly, the lattice order parameter will also start oscillating after photo--excitation, and
its real--time evolution can be calcualted by using Eq.~\eqref{eq:real_time}.
In order to illustrate the lattice response, we focus on the $\gS$ valley and assume that the photo--excitation occurs at time 
$T=T_\mathrm{B}$ (see \fig{fig:order_parameter_T}).
The resulting phonon displacement time dependence is shown in \fig{fig:phonon_osc} and reveals that the lattice will start oscillating
as a consequence of the symmetry breaking DEI$\rar$IEI transition. This oscillation could be observed, for example, in
a time--dependent transient absorption~\cite{Merlin1997,Kuznetsov1994} or crystallography~\cite{KVICK2017648} experiment following
the probe photo--excitation.

\mysec{In conclusion}
We present a joint experimental and theoretical description of the ultrafast valley dynamics observed in the \trarpes spectrum of WSe$_2$.  The experimentally
observed transfer of spectral weight from the $K$ to the $\gS$ valley is interpreted as a photo--induced transition from a direct to an
indirect excitonic--insulating phase.
The present description is not bound to the linear regime and, thus, does not require the introduction of an excitonic quasiparticle picture.
In order to numerically tackle the problem we introduce a non--equilibrium time
separation approach where slowly evolving incoherent carrier dynamics triggers the formation of an adiabatic excitonic insulating phase that moves
from the $K$ to the $\gS$ valley, inducing a \trarpes dynamics in excellent agreement with the experimental results.
We also predict the appearance of a time--dependent lattice order parameter that, in agreement with the Rice and Kohn prediction~\cite{JeromeRiceKohn1967}, 
ensures total momentum conservation. The lattice,
indeed, starts moving opposite to the electronic density matrix.
Our results provide an entirely coherent, well-defined and sound interpretation of the \trarpes dynamics observed experimentally. Rather than using
an excitonic quasiparticle picture, our method relies on the photo--induced dynamics of an oscillating order parameter. Besides explaining the experimental
results, we also predict the appearance of coherent lattice oscillations triggered by the photo--excitation. This opens the possibility of observing
further experimental evidence of our proposed finite-momentum photo--excited excitonic insulating phase.

\bibliography{reference,EI}

\end{document}